\documentclass[twocolumn,secnumarabic,amssymb, nobibnotes, aps, prb]{revtex4-1}

\usepackage{titlesec}
\titlespacing*{\section}
{0pt}{2.5ex plus 1ex minus .1ex}{1.3ex plus .1ex}
\titleformat{\section}
{\normalfont\fontsize{10}{10}\bfseries\filcenter}{\thesection}{1em}{}
\usepackage{amsfonts}
\usepackage{amsmath}
\usepackage{amssymb}
\usepackage{graphicx}
\usepackage{epstopdf}
\usepackage{comment}
\usepackage[toc,page]{appendix}
\usepackage{float}%
\begin{document}
		\begin{abstract}
The momentum transport in ultraclean bilayer graphene is characterized by the viscous transport. In quantizing magnetic field the momentum current passes through the guiding centers of cyclotron orbits. In this study we derive the formula of quantized Hall viscosity of bilayer graphene that is the next topological feature after the quantum Hall effect. This can be detected in the non-local magnetoresistivity measurements that varies with the quantized steps of magnetoresistivity.
		\begin{description}
			\item[Usage]
			.
			\item[PACS numbers]
			May be entered using the \verb+\pacs{#1}+ command.
			\item[Structure]
			You may use the \texttt{description} environment to structure your abstract;
			use the optional argument of the \verb+\item+ command to give the category of each item. 
		\end{description}
	\end{abstract}
	\title{Quantizing momentum transport in bilayer graphene}%
	
	\author{Muhammad Imran}%
	\affiliation{Department of Physics, University of Florida, Gainesville, Florida 32611,USA}
	\date{\today}%
	\maketitle
		\subsubsection{INTRODUCTION}
		Graphene is a very famous material that host exotic transport properties. However an ultraclean graphene is used as a platform for momentum transport.\cite{Ref:13} Quite recently the momentum transport in graphene has become a hot topic of interest.\cite{Ref:4} This has become possible because graphene possess ultrahigh mobility and allows both ballistic and diffusive transport. In ballistic transport region the particles mean free path length, $l_{M}\sim 1\mu m$,\cite{Ref:16} is greater than the device size $l_{S}$, $\frac{{l_{M}}}{l_{S}}>1$, here particles transport is device size dependent. In the ultraclean but collision dominated region interparticles scattering length $l_{MC}\sim 1\mu m$,\cite{Ref:16} is smaller than the device size, $\frac{l_{MC}}{l_{S}}<<1$, the particles transport is diffusive and follow Poiseuille flow. The scattering processes in a dirty sample conserve the number of particles but degrades the momentum. This makes charge transport ohmic. In the ultraclean sample the momentum is also conserved and the charge transport is diffusive. However in the presence of strong magnetic field $B>10 T$ the time reversal symmetry is broken and the motion of charge particles is quantized. The charge transport in the direction perpendicular to magnetic and electric fields is fault tolerant. The ultraclean samples not only allow quantization of charge transport but also the quantization of momentum transport.\cite{Ref:17} The voltage induced by Hall viscosity is opposite in the direction as compare to the Hall conductivity. This viscous Hall voltage has already been observed in the graphene.\cite{Ref:11}
		
		The semiclassical theory of viscous transport in electron gas was first developed by Gurzhi and his coworkers. \cite{Ref:26,Ref:27,Ref:28} Later the formula of Hall viscosity was derived that depends on scattering time.\cite{Ref:18,Ref:24} Experimentally viscous transport has been observed in $GaAs$,\cite{Ref:22, Ref:19,Ref:20,Ref:21,Ref:39} Graphene,\cite{Ref:11} $PdCoO_{2}$,\cite{Ref:37} and $WP_{2}$.\cite{Ref:38} The ultraclean sample of $GaAs$ has mobility $\mu\sim10^{7}\frac{cm^{2}}{Vs}$ at low temperature and therefore allows both viscous and ballistic transport. The temperature dependent negative magnetoresistance in $GaAs$ is explained by the viscous transport.\cite{Ref:25, Ref:19,Ref:20,Ref:21,Ref:22,Ref:23,Ref:24} This is an interesting fact that temperature dependence of negative magnetoresistance in $GaAs$ survives even when interparticles scattering length is greater than device size, $\frac{l_{MC}}{l_{S}}>1$.\cite{Ref:23} In the presence of time reversal symmetry this should depend on the sample size being in the ballistic region of charges transport.
		
		The quantum theory of viscous transport does not require a scattering time in the absence of time reversal or/and inversion symmetries.\cite{Ref:6} The quantized Hall viscosity formula is derived by straining the metric of electron gas.\cite{Ref:2,Ref:6, Ref:32,Ref:33,Ref:34} However there is no experimental evidence of the observation of quantized Hall viscosity.

\subsubsection{Model}
		In this work we study the momentum transport of particles and holes in quantizing magnetic field applied to bilayer graphene. This has been done for two dimensional electron gas and monolayer graphene.\cite{Ref:2,Ref:3,Ref:5} Graphene is very rich in physics due to gapless energy spectrum and allows both particles and holes transport. The momentum transport of particles and holes involves viscous response of the system. Graphene is very viscous semimetal with its viscosity exceeding the viscosity of honey by an order of magnitude $\nu\sim 0.1\frac{m^{2}}{s}$.\cite{Ref:13,Ref:14} We derive the formulas of viscous transport by considering strain as a perturbation applied to the metric of bilayer graphene. The response of strain as a perturbation for the formula of stress-stress correlation function is the same as the response of magnetic vector potential for the formula of current-current correlation function.
		The effective Hamiltonian of bilayer graphene in the quantizing magnetic field is,\cite{Ref:7} 
		\begin{equation}
		H=\frac{\Pi_{x}^{2}-\Pi_{y}^{2}}{2m}\sigma_{x}+\frac{\Pi_{x}\Pi_{y}+\Pi_{y}\Pi_{x}}{2m}\sigma_{y}.
		\end{equation}
Here $\vec{\Pi}=\vec{p}-e\vec{A}(r)$ is the kinetic momentum, $\vec{A}(r)$ is vector potential, and $m$ is the band mass of charge particle. $\vec{\sigma}$ is Pauli matrix and denotes the valley-orbit coupling. The corresponding energy eigenvalues and energy eigenstates are,
\begin{eqnarray}
\psi_{m}(x,y)&=&\frac{Exp(i\vec{k}\cdot\vec{x})}{\sqrt{2\mathcal{A}}}\begin{bmatrix}
\phi_{m}(y+y_{0})  \\
r\phi_{m-2}(y+y_{0})\\
\end{bmatrix}\\
\epsilon_{m}^{r}&=&r\hbar\omega_{c}\sqrt{m(m-1)}.
\end{eqnarray}
Here $\phi_{m}(y+y_{0})$ are eigenstates of the Harmonic oscillator and $r=\pm$. $\omega_{c}$ is cyclotron frequency, $y_{0}$ is the shift in Harmonic oscillator due to vector potential and momentum coupling, and $\mathcal{A}$ is the area of the sample.

The stress tensor $T_{\mu\nu}$ for two dimensional fluid is proportional to the velocity gradient $v_{\alpha\beta}=\partial_{\alpha}v_{\beta}+\partial_{\beta}v_{\alpha}$ with proportionality constant viscosity $\nu_{\mu\nu\alpha\beta}$. 

\begin{equation}
T_{\mu\nu}=\nu_{\mu\nu\alpha\beta}v_{\alpha\beta}
\end{equation}
Here Einstein summation convention is assumed in the repeated indexes.
The viscosity for two dimensional electron gas has sixteen components,
\begin{equation}
\begin{split}
\nu_{\mu\nu\alpha\beta}=\nu_{xxxx}\delta_{\mu,\nu}\delta_{\alpha,\beta}+\nu_{xyxy}(\delta_{\mu,\alpha}\delta_{\nu,\beta}+\delta_{\mu,\beta}\delta_{\nu,\alpha}\\
-\delta_{\mu,\nu}\delta_{\alpha,\beta})+\frac{1}{2}\nu_{xxxy}(\epsilon_{\mu\alpha}\delta_{\nu\beta}+\epsilon_{\mu\beta}\delta_{\nu\alpha}+\epsilon_{\nu\alpha}\delta_{\mu\beta}\\
+\epsilon_{\nu\beta}\delta_{\mu\alpha}),
\end{split}
\end{equation}  
but the symmetry argument narrows down these possibilities into only three components, the symmetric components of viscosity $\nu_{xxxx}$, $\nu_{xyxy}$, and the antisymmetric component of viscosity(Hall viscosity) $\nu_{xxxy}$.
These components of the viscosities are derived from the traceless stress tensor operator. The traceless stress tensor operator of bilayer graphene is constructed by following the same steps as is needed for ordinary two dimensional electron gas, except the pseudospin degree of freedom is added.\cite{Ref:2} The stress tensor operator different components are found by using this formula,
\begin{equation}
T_{\mu\nu}=-\frac{i}{\hbar}[H,J_{\mu\nu}^{T}],
\end{equation}
where this commutation relation is derived by considering a linear order correction of strain perturbation in unstrained Hamiltonian.\cite{Ref:2}
Here $J_{\mu\nu}^{T}$ is the total traceless stress generator of bilayer graphene, $J_{\mu\nu}^{T}=L_{\mu\nu}+S_{\mu\nu}$, and this is the combination of the orbital angular momentum $L_{\mu\nu}$ and pseudospin angular momentum $S_{\mu\nu}$. The angular and spin components of the traceless stress generator of bilayer graphene are,\cite{Ref:2}
\begin{eqnarray}
L_{\mu\nu}&=&-\frac{1}{2}\{x_{\mu},\Pi_{\nu}\}+\frac{1}{4}\{x_{\alpha},\Pi_{\alpha}\}\delta_{\mu,\nu}+\frac{eB}{2}\epsilon_{\nu\alpha}x_{\mu}x_{\alpha}\\
S_{\mu\nu}&=&-\hbar\sigma_{z}\\
0&=&[H,J_{\mu\nu}^{Tz}].
\end{eqnarray}
Here $J_{\mu\nu}^{Tz}$ is the conserved component of the total angular momentum that ensures a symmetric traceless stress tensor operator. The above formula is made consistent with valley-orbit coupling by exploiting conservation of total momentum.\cite{Ref:1} The longitudinal and transverse stress tensor operators of bilayer graphene are,
\begin{eqnarray}
T_{\mu\nu}&=&\frac{\Pi_{x}^{2}+\Pi_{y}^{2}}{2m}(\sigma_{x}\times\tau_{\mu\nu}^{z}+\sigma_{y}\times\tau_{\mu\nu}^{x}).
\end{eqnarray} 
Here $\tau^{x(z)}_{\mu\nu}$ are Pauli spinors and used to write the different components of the stress tensor operator. The Kubo linear response theory gives the longitudinal and Hall viscosities of bilayer graphene.
The stress-stress response function is,
\begin{equation}
Q_{\mu\nu\alpha\beta}(\Omega)=g\frac{k_{B}T}{2\pi l_{B}^{2}}\sum_{m,n,\omega}Tr[T_{\mu\nu}G(\Omega+\omega,\xi_{m}^{r})T_{\alpha\beta}G(\omega,\xi_{n}^{s})].
\end{equation}
The Greens function is,
\begin{equation}
\begin{split}
G(x,x',\omega,\xi_{m}^{r})=\frac{g(\omega,\xi_{m}^{r})}{2}\\
\begin{bmatrix}
\phi_{m}(Y)\phi_{m}(Y')  && r\phi_{m}(Y)\phi_{m-2}(Y')\\
r\phi_{m-2}(Y)\phi_{m}(Y')  && \phi_{m-2}(Y)\phi_{m-2}(Y')
\end{bmatrix},
\end{split}
\end{equation}
where $g(\omega,\xi_{m}^{r})=(\omega-\xi_{m}^{r}+\frac{i}{2\tau}sgn(\omega))^{-1}$, $\xi_{m}^{r}=\epsilon_{m}^{r}-\mu$, $Y=y+y_{0}$. $sgn(\omega)=1$ for $\omega>0$ and $sgn(\omega)=-1$ for $\omega<0$.  $\mu$ denotes chemical potential. $\omega$ and $\Omega$ are matsubara frequencies. $g$ denotes the degeneracy of the bilayer graphene. In the above formula the diamagnetic response is not included. We will include it in the formula of viscosity. 
\begin{equation}
\begin{split}
\nu_{xxxx}=\frac{g\hbar\omega_{c}}{2\pi l_{B}^{2}\bar{m}}\delta(\omega)\sum_{m}\sqrt{m(m-1)}\\
+\frac{g\hbar}{4\pi l_{B}^{2}\bar{m}}\frac{\beta\hbar}{\tau}\sum_{m}(m+\frac{1}{2})^{2}F^{+,+}_{m,m+2}
\end{split}
\end{equation} 	
\begin{equation}
\begin{split}
\nu_{xxxy}=\frac{g\hbar}{\pi l_{B}^{2}\bar{m}}\sum_{m}[(m^{2}+m+1)G_{m,m+2}
\end{split}
\end{equation}
\begin{equation}
\begin{split}
F^{\alpha,\gamma}_{m,n}=[\frac{\hbar^{2}\omega_{c}^{2}}{(\epsilon_{m}-\epsilon_{n})^{2}+\frac{\hbar^{2}}{\tau^{2}}}\\
+\alpha\frac{\hbar^{2}\omega_{c}^{2}}{(\epsilon_{m}+\epsilon_{n})^{2}+\frac{\hbar^{2}}{\tau^{2}}}]\\
[sech^{2}[\frac{\beta\xi_{m}^{+}}{2}]+sech^{2}[\frac{\beta\xi_{m}^{-}}{2}]\\+\gamma(sech^{2}[\frac{\beta\xi_{n}^{+}}{2}]+sech^{2}[\frac{\beta\xi_{n}^{-}}{2}])]
\end{split}
\end{equation}
\begin{equation}
\begin{split}
G_{m,n}=\tanh[\frac{\beta\xi_{m}^{+}}{2}]+\tanh[\frac{\beta\xi_{m}^{-}}{2}]\\
-\tanh[\frac{\beta\xi_{n}^{+}}{2}]-\tanh[\frac{\beta\xi_{n}^{-}}{2}]
\end{split}
\end{equation}
Here $\nu_{xxxx}=\nu_{xyxy}$ and $\bar{m}$ is the mass density.
For the longitudinal viscosity we assume a finite lifetime $\tau$ of quasiparticles. This should be reminded that viscosity is also subjected to the holographic bound to restrict the transport time, $\frac{\bar{m}\nu}{s}\geq\frac{h}{2k_{B}}$ (where $\nu$ is viscosity and $s$ is entropy density).\cite{Ref:30} However this inequality does not hold for quantized Hall viscosity,\cite{Ref:24} and in the anisotropic liquids.\cite{Ref:40} The Hall viscosity shows the quadrupolar excitations and only survives in the absence of time reversal symmetry. 

The viscous response of charged liquid is observed either in the Hall bar geometry\cite{Ref:11, Ref:14} or in the Carbino disk geometry.\cite{Ref:35} In either case the non-local current density and applied source equation is solved to find the viscous response. For bilayer graphene we solve the non-local current and field equation in the Hall bar geometry,
\begin{equation}
\nu_{xxxx}\nabla^{2}\vec{J}+\nu_{xxxy}(\hat{z}\times\nabla^{2}\vec{J})=\frac{(ne)^{2}}{\bar{m}}\vec{\nabla}\phi(x,y).
\end{equation}
The longitudinal viscosity is a dissipative response and in pristine samples the parramagnetic response of the longitudinal viscosity is negligible compare to Hall viscosity. $\phi(x,y)$ is induced potential due to the viscosity. This potential accumulates at the location of applied source or sink. Here we consider a half plane geometry. The induced potential is derived from the equation of stream function, $\psi(x,y)$, with its curl equals to the current density, $\vec{J}=\hat{z}\times\vec{\nabla}\psi$.
\begin{equation}
\begin{split}
\nu_{xxxx}\nabla^{2}(\hat{z}\times\vec{\nabla}\psi(x,y))-\nu_{xxxy}\nabla^{2}(\vec{\nabla}\psi(x,y))\\
=\frac{(ne)^{2}}{\bar{m}}\vec{\nabla}\phi(x,y)
\end{split}
\end{equation} 

We consider a point current source, $J_{y}(x,0)=I\delta(x)$, with no stress boundary conditions, $\partial_{y}^{2}\psi(x,y)|_{y=0}=0$. The induced potential is,
\begin{equation}
\phi(x,y)=\frac{\bar{m}I}{(ne)^{2}\pi}\frac{\nu_{xxxx}(x^{2}-y^{2})-2\nu_{xxxy}xy}{(x^{2}+y^{2})^{2}}.
\end{equation}
This induced potential increases in discrete steps by increasing magnetic field, since Hall viscosity is quantized.

\subsubsection{The ohmic and viscous magnetotransport }The non-local current field relation also involves the ohmic contribution.\cite{Ref:9} This motivates us to derive the nonlocal magnetoresistance formula from the current-current correlation function.
The current-current response function is,\cite{Ref:12}
\begin{equation}
\begin{split}
Q_{\mu\nu}(\Omega)=g\frac{k_{B}T}{2\pi l_{B}^{2}}\sum_{m,n,\omega}Tr[v_{\mu}\exp[i\bar{q}\bar{\Pi}_{x}]G(\Omega+\omega,\xi_{m}^{r})\\
\exp[-i\bar{q}\bar{\Pi}_{x}]v_{\nu}G(\omega,\xi_{n}^{s})].
\end{split}
\end{equation}
where $v_{x}=\frac{\Pi_{x}}{m}\sigma_{x}+\frac{\Pi_{y}}{m}\sigma_{y}$, and $v_{y}=-\frac{\Pi_{y}}{m}\sigma_{x}+\frac{\Pi_{x}}{m}\sigma_{y}$ are velocity operators. $\bar{q}=ql_{B}$ is dimensionless wavevector and $\exp[\pm i\bar{q}\bar{\Pi}_{x}]$ is the shift operator.\cite{Ref:3, Ref:36}

The current-current correlation function is expanded upto second order in the dimensionless wavevector $O(\bar{q})^{2}$. The longitudinal and Hall conductivities are,
\begin{equation}
\begin{split}
\sigma_{xx}=\sigma_{l}(\sigma_{xx}^{(0)}+\bar{q}^{2}\sigma_{xx}^{(2)})\\
\sigma_{xy}=\sigma_{t}(\sigma_{xy}^{(0)}+\bar{q}^{2}\sigma_{xy}^{(2)}),
\end{split}
\end{equation}
where $\sigma_{l}=g\frac{e^{2}}{8\pi\tau k_{B}T}$, and $\sigma_{t}=g\frac{e^{2}}{h}$. 
\begin{equation}
\sigma_{xx}^{(0)}=\sum_{m}[mF^{+,+}_{m,m+1}]
\end{equation}
\begin{equation}
\begin{split}
\sigma_{xy}^{(0)}=\sum_{m}[mG_{m,m+1}]
\end{split}
\end{equation}
\begin{equation}
\begin{split}
\sigma_{xx}^{(2)}=2\sum_{m}[m^{2}(F_{m,m+2}^{+,+}-2F_{m,m+1}^{+,+})\\
+(2m+1)F_{m,m+1}^{+,0}-m\sqrt{m^{2}-1}F_{m,m+1}^{-,+}\\
-mF_{m,m+1}^{+,-}+2m(m-1)g_{m}]
\end{split}
\end{equation}

\begin{equation}
\begin{split}
\sigma_{xy}^{(2)}=\sum_{m}[\frac{m^{2}(m^{2}+m+1)}{8(m+\frac{1}{2})^{2}}G_{m,m+2}\\
-m^{2}G_{m,m+1}]
\end{split}
\end{equation}

\begin{equation}
\begin{split}
g_{m}=\omega_{c}^{2}\tau^{2}[sech^{2}[\frac{\beta\xi_{m}^{+}}{2}]+sech^{2}[\frac{\beta\xi_{m}^{-}}{2}]].
\end{split}
\end{equation}
The second order expansion of magnetoconductivity in wavevector $\bar{q}$ is proportional to the Hall viscosity(the other terms are also present in an ordinary two dimensional electron gas and monolayer graphene). To include the effect of point current source and find its response as the induced potential we write the Navier Stokes equation, derived from the non-local field and magnetoconductivity relation. The point current source response in nonlocal magnetoresistance is derived by converting $\bar{q}\rightarrow l_{B}\vec{\nabla}$, \cite{Ref:3,Ref:5}
\begin{equation}
-\vec{\nabla}\phi=(\rho_{xx}^{(0)}+\rho_{xx}^{(2)}l_{B}^{2}\Delta)\vec{J}+(\rho_{xy}^{(0)}+\rho_{xy}^{(2)}l_{B}^{2}\Delta)\vec{J}\times\hat{z},
\end{equation} 
where $\rho_{xx}^{(0,2)}$ and $\rho_{xy}^{(0,2)}$ are the magnetoresistivity matrix elements. $l_{B}$ is magnetic length and $\Delta$ is the Laplacian operator. The above equation is broader in a sense that it deals with both the ohmic($\rho^{(0)}_{xx},\rho^{(0)}_{xy}$) and viscous($\rho^{(2)}_{xx},\rho^{(2)}_{xy}$) transport parameters,\cite{Ref:9,Ref:10} whereas Eq: 18 only deals with the viscous transport parameters($\nu_{xxxx},\nu_{xxxy}$). 

The same setup is solved for an ordinary two dimensional electron gas in Ref[8]. We derive the formula of non-local resistivity $R_{nl}$ by solving above equation. For this we use the no stress boundary condition $\partial^{2}_{y}\psi_{k}(y)|_{y=0}=0$ and consider a Lorentzian current source $J_{y}(x,0)=\frac{1}{\pi}\frac{I\delta}{\delta^{2}+x^{2}}$. The non-local resistance is,
\begin{equation}
\begin{split}
R_{nl}(|x|)=\frac{\rho_{xy}^{0}}{2}(1-4l_{B}^{2}|\Gamma_{H}|\frac{|x|\delta}{(x^{2}+\delta^{2})^{2}}),
\end{split}
\end{equation} 
where $\delta$ is the finite width of the point contact and $\Gamma_{H}=\frac{\rho_{xy}^{(2)}}{\rho_{xy}^{(0)}}$. The nonlocal magnetoresistance is derived by considering zero potential at the opposite side of the sample $R_{nl}(|x|)=\frac{\phi(|x|,0)-\phi(|x|,\infty)}{I}$.\cite{Ref:9,Ref:10} The nonlocal magnetoresistance becomes negative close to the point contact, a signature of fluidity.\cite{Ref:9} By this formula we see this nonlocal magnetoresistance should vary in discrete steps, since magnetoresistivity is quantized.    
\subsubsection{Discussion}
The quantized Hall and longitudinal viscosities are plotted in Fig. 1. The functional dependence of Hall and longitudinal viscosities in the chemical potential is identical with the Hall and longitudinal magnetoresistivities, except the viscosities take two steps instead of one step at any time for being a quadrupolar response of applied perturbation. This viscous response is probed in the magnetotransport experiment. In the presence of point current source we see in Fig. 2 the induced potential changes it sign with the spatial variation. The stream function at the close vicinity of the current source takes a U turn. These features are the properties of the viscous transport in bilayer graphene. In this study we also emphasize the quantized variation in the induced potential with Hall viscosity. This is plausible in strong magnetic field for induced potential to stay constant unless the next Landau level enters the conduction channel by tuning magnetic field. Fig .2 (b), (c), and (d) shows the induced potential profile depends on the strength of the ratio of the longitudinal viscosity to the Hall viscosity. This shows an asymmetric potential profile arises in the presence of magnetic field, since the Hall viscosity becomes finite when magnetic field is present. The potential induced by the Hall viscosity have different spatial dependence than the longitudinal viscosity.  

 \begin{figure}[t!]
	\centering
	\includegraphics[width=90mm,height=65mm]{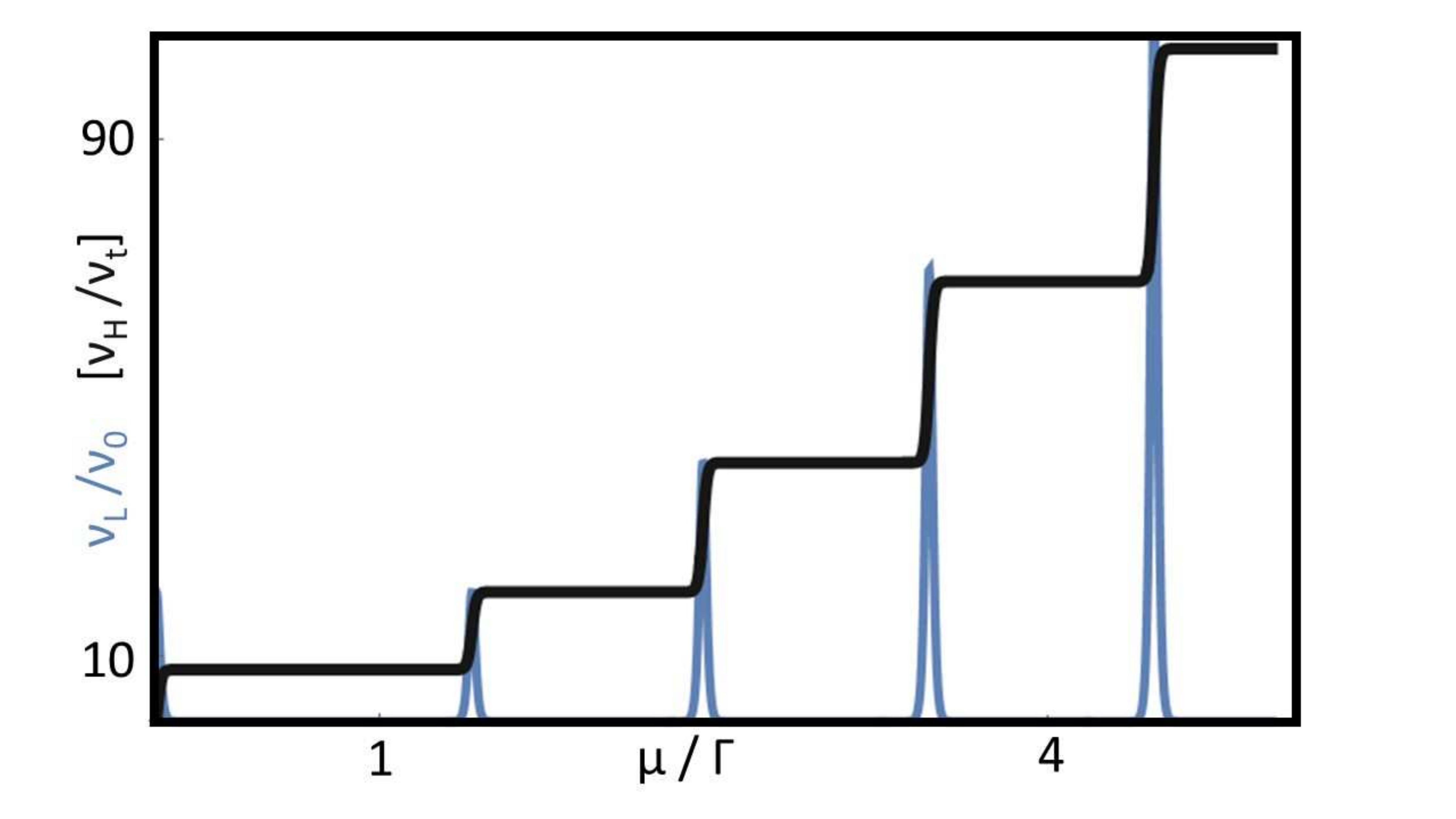}
	\caption{The quantized Hall and longitudinal viscosities. Here $\frac{\nu_{H}}{\nu_{t}}=\frac{\pi l_{B}^{2}\bar{m}\nu_{xxxy}}{gh}$ and $\frac{\nu_{L}}{\nu_{0}}=\beta\Gamma\frac{4\pi l_{B}^{2}\bar{m}\nu_{xxxx}}{gh}$ and this does not include the diamagnetic response. $\beta\Gamma=10$ and $\frac{\hbar\omega_{c}}{\Gamma}=10$. \label{fig:figgg}}
\end{figure}
\begin{figure}[t!]
	\centering
	\includegraphics[width=90mm,height=65mm]{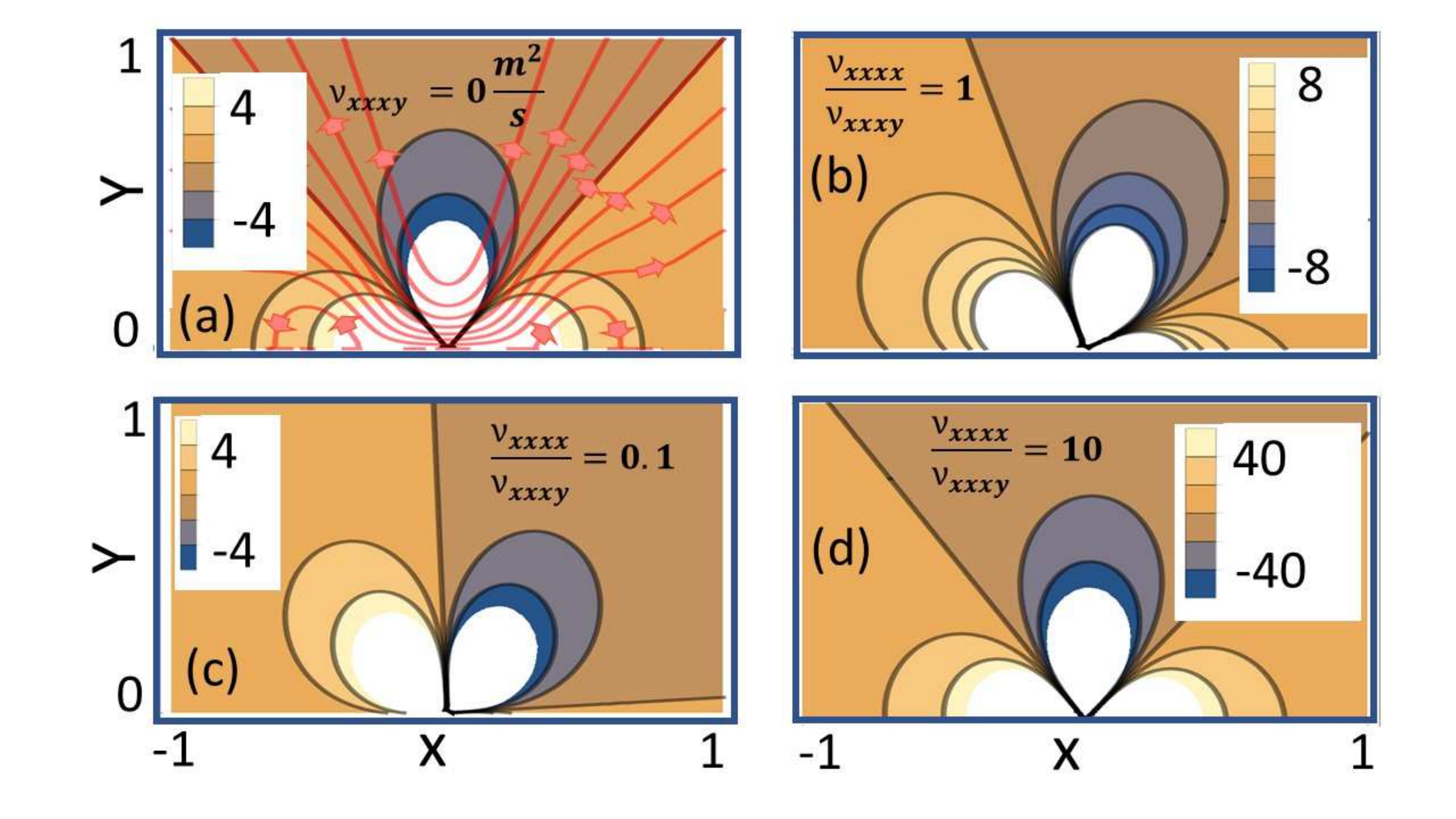}
	\caption{Contour plot of induced potential and stream function. Here $X=\frac{x}{l_{B}}$ and $Y=\frac{y}{l_{B}}$. In Fig. (a) stream function is shown by the red lines with arrows on top of it and Hall viscosity is zero. Figs. (b), (c), and (d) show induced potential profile for different ratios of the longitudinal and Hall viscosities.  \label{fig:figgg}}
\end{figure}
\begin{figure}[b!]
	\centering
	\includegraphics[width=90mm,height=55mm]{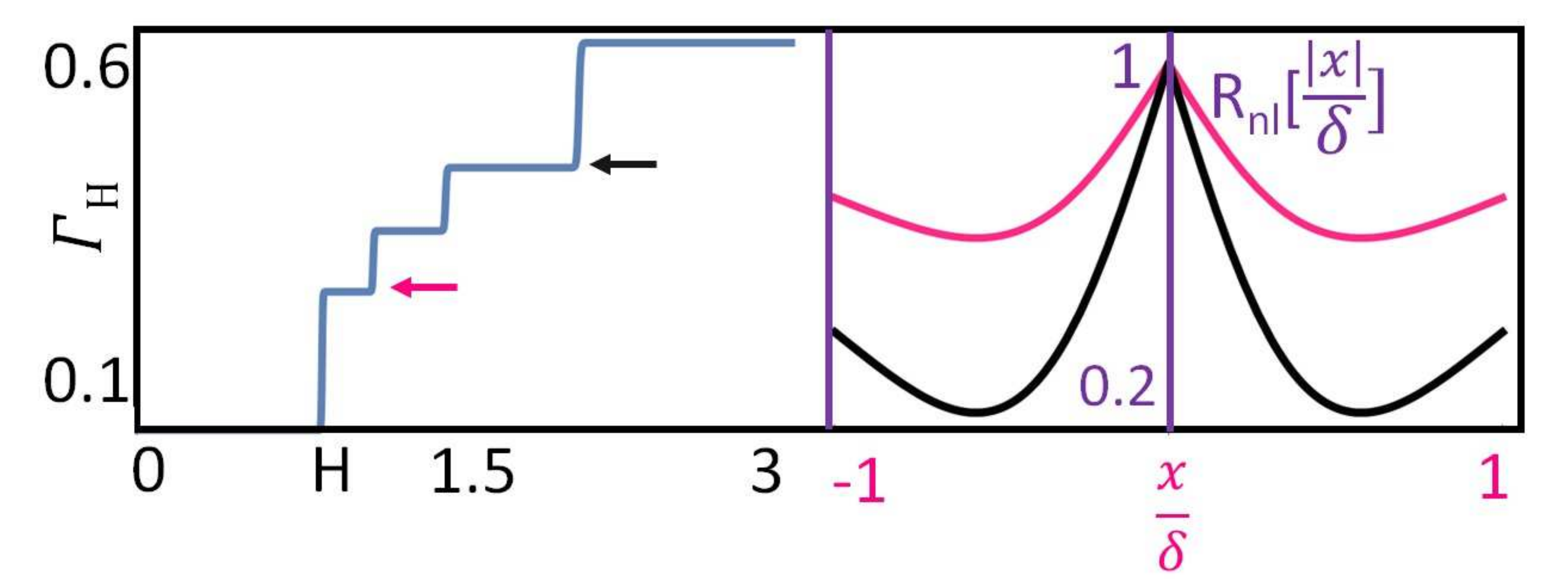}
	\caption{Nonlocal magnetoresistance, $R_{nl}$, for $\Gamma_{H}=0.2$, $\Gamma_{H}=0.4$, $l_{B}=2.5\delta$, $H=\beta\hbar \omega_{c}$, and $\beta\mu=500$.  \label{fig:figgg}}
\end{figure}
A direct relation between the Hall and longitudinal viscosities exists with the Hall and longitudinal magnetoconductivities. This is proved by expanding the magnetoconductivities upto second order in the dimensionless wavevector and later we have used it to derive the Navier Stokes equation. The induced potential is solved for a point current source. Another observation is the negative sign of the induced potential than the ordinary Hall signal. This is an indication of the viscous transport where ordinary Hall voltage is against the direction of the viscous flow. In Fig. 3 we plot the non-local resistance in the vicinity of an applied current source. This non-local magnetoresistance decreases with moving away from the point source. By tuning the external magnetic field the profile of the non-local magnetoresistance changes in a quantized manner. This can be the smoking gun evidence for the observation of quantized Hall viscosity.    
 
The experimental observation of quantized Hall viscosity is an open question. Quite recently Hall viscosity is measured in graphene samples of high quality, with mobility as high as $\mu=10^{5}\frac{cm^{2}}{Vs}$, that survives upto room temperature.\cite{Ref:11} However these measurements are made in weak magnetic field, $B\sim 10mT$, to keep cyclotron orbits far greater than the device size. The observation of quantized Hall viscosity is possible in magnetic field $B\geq10T$. This range of magnetic field is four order of magnitudes greater than the present data on Hall viscosity measurements. Therefore the measurements of quantized Hall viscosity is a daunting challenge from the experimental side.

\subsubsection{Conclusion}
In this study we have derived the formula of quantized Hall viscosity in bilayer graphene. The quantized viscous transport is expected to create quantized variation in the magnitude of induced potential. We have also derived the Navier Stokes equation that predicts quantized viscous transport in non-local magnetoresistance. The leading order formula of non-local magnetoresistance does not depend on dissipative components of the magnetoresistivity and viscosity. This makes this quantity a fault tolerant feature of the magnetotransport that does not modify due to the scattering. This quantized momentum transport in perpendicular direction of applied electric and magnetic fields is the next topological feature than the quantized charge transport. Its sign is opposite to the charge transport and therefore creates whirlpool in the vicinity of the current source. The quantized momentum transport will be another promising candidate for the fault tolerant information transport via momentum, if experimentally observed.


\begin{thebibliography}{99} 
		\bibitem{Ref:1}Julia M. Link, Daniel E. Sheehy, Boris N. Narozhny, and Jörg Schmalian, Phys. Rev. B \textbf{98}, 195103 (2018).
		\bibitem{Ref:2}Barry Bradlyn, Moshe Goldstein, and N. Read, Phys. Rev. B \textbf{86}, 245309 (2012).
		\bibitem{Ref:3}Mohammad Sherafati, Alessandro Principi, and Giovanni Vignale, Phys. Rev. B \textbf{94}, 125427 (2016).
		\bibitem{Ref:4}Y. Nam, D.-K. Ki, D. Soler-Delgado, and A. F. Morpurgo, Nat. Phys. 13, 1207 (2017).
		\bibitem{Ref:5}Carlos Hoyos and Dam Thanh Son, Phys. Rev. Lett. 108,
		066805 (2012).    
	    \bibitem{Ref:6}J. Avron, R. Seiler, and P. G. Zograf, Phys. Rev. Lett. 75, 697 (1995).
	    \bibitem{Ref:7}Edward McCann and Vladimir I. Fal’ko, Phys. Rev. Lett. 96, 086805 (2006). 
	    \bibitem{Ref:8}Luca V. Delacretaz and Andrey Gromov, Phys. Rev. Lett. 119, 226602 (2017). 
	    \bibitem{Ref:9}L. Levitov and G. Falkovich, Nat. Phys. 12, 672 (2016).
	    \bibitem{Ref:10}G. Falkovich and L. Levitov, Phys. Rev. Lett. 119, 066601 (2017).
	    \bibitem{Ref:11}A. I. Berdyugin, S. G. Xu, F. M. D. Pellegrino, R. Krishna Kumar, A. Principi, I. Torre, M. Ben Shalom, T. Taniguchi, K. Watanabe, I. V. Grigorieva, M. Polini, A. K. Geim, and D. A. Bandurin, Science 364, 162 (2019).
	    \bibitem{Ref:12}The diamagnetic response is not included in this derivation. The longitudinal conductivity diverges for $\Omega=0$. However the non-local magnetoresistance turns out to be independent of longitudinal response of current-current response function.
	    \bibitem{Ref:13}A. Lucas, Science 364, 6436(125) (2019).
	    \bibitem{Ref:14}D. A. Bandurin, I. Torre, R. Krishna Kumar, M. Ben Shalom, A. Tomadin,
	    A. Principi, G. H. Auton, E. Khestanova, K. S. Novoselov, I. V. Grigorieva,
	    L. A. Ponomarenko, A. K. Geim, M. Polini, Science 351, 1055 (2016).
	    \bibitem{Ref:15}I. Torre, A. Tomadin, A. K. Geim, and M. Polini, Phys. Rev. B 92, 165433 (2015).
	    \bibitem{Ref:16}R. Krishna Kumar, D. A. Bandurin, F. M. D. Pellegrino, Y. Cao, A. Principi, H. Guo, G. H. Auton, M. Ben Shalom, L. A. Ponomarenko, G. Falkovich, K. Watanabe, T. Taniguchi, I. V. Grigorieva, L. S. Levitov, M. Polini and A. K. Geim, Nat. Phys. 13, 1182 (2017).
	    \bibitem{Ref:17}J. Avron, J. Stat. Phys. 92, 543 (1998).
	    \bibitem{Ref:18}M. S. Steinberg, Phys. Rev. 109, 1486 (1958).
	   \bibitem{Ref:19} A. T. Hatke, M. A. Zudov, J. L. Reno, L. N. Pfeiffer, and
	    K.W. West, Phys. Rev. B 85, 081304 (2012).
	    \bibitem{Ref:20}R. G. Mani, A. Kriisa, and W. Wegscheider, Scientific
	    reports 3, 2747 (2013).
	    \bibitem{Ref:21}L. Bockhorn, P. Barthold, D. Schuh, W. Wegscheider, and
	    R. J. Haug, Phys. Rev. B 83, 113301 (2011)
	    \bibitem{Ref:22}L. Bockhorn, A. Hodaei, D. Schuh, W. Wegscheider, and R. J. Haug,
	    J. Phys. Conf. Ser. 456, 012003 (2013).
	    \bibitem{Ref:23}P. S. Alekseev, Phys. Rev. Lett. 117, 166601 (2016).
	    \bibitem{Ref:24}T. Scaffidi, N. Nandi, B. Schmidt, A. P. Mackenzie, and J. E. Moore, Phys. Rev. Lett. 118, 226601 (2017).
	    \bibitem{Ref:25}Q. Shi, P. D. Martin, Q. A. Ebner, M. A. Zudov, L. N.
	    Pfeiffer, and K.W. West, Phys. Rev. B 89, 201301 (2014).
	    \bibitem{Ref:26}R. N. Gurzhi, Sov. Phys. JETP 17, 521 (1963).
	    \bibitem{Ref:27}R. N. Gurzhi and S. I. Shevchenko, Sov. Phys. JETP 27,
	    1019 (1968).
	    \bibitem{Ref:28}R. N. Gurzhi, A. N. Kalinenko, and A. I. Kopeliovich, Phys.
	    Rev. Lett. 74, 3872 (1995).
	    \bibitem{Ref:30}M. Müller, J. Schmalian, and L. Fritz, Phys. Rev. Lett. 103, 025301 (2009)
	    \bibitem{Ref:31}N. Read, Phys. Rev. B 79, 045308 (2009).
	    \bibitem{Ref:32}N. Read and E. H. Rezayi, Phys. Rev. B 84, 085316
	    (2011).
	    \bibitem{Ref:33}T. L. Hughes, R. G. Leigh, and E. Fradkin, Phys. Rev. Lett.
	    107, 075502 (2011).
	    \bibitem{Ref:34}T. L. Hughes, R. G. Leigh, and O. Parrikar, Phys. Rev. D 88,
	    025040 (2013).
	    \bibitem{Ref:35}A. Tomadin, G. Vignale, and M. Polini, Phys. Rev. Lett. 113, 235901 (2014).
	    \bibitem{Ref:36}F. D. M. Haldane, arXiv:0906.1854.
	    \bibitem{Ref:37}P. J. W. Moll, P. Kushwaha, N. Nandi, B. Schmidt, and A. P. Mackenzie, Science 351, 1061 (2016).
	    \bibitem{Ref:38}J. Gooth, F. Menges, N. Kumar, V. Suß, C. Shekhar, Y. Sun,
	    U. Drechsler, R. Zierold, C. Felser, and B. Gotsmann, Nat.
	    Commun. 9, 4093 (2018).
	    \bibitem{Ref:39}B. A. Braem, F. M. D. Pellegrino, A. Principi, M. Roosli, C. Gold, S. Hennel, J. V. Koski, M. Berl, W. Dietsche,
	    W. Wegscheider, M. Polini, T. Ihn, and K. Ensslin, Phys. Rev. B 98, 241304(R)
	    (2019).
	    \bibitem{Ref:40}J. M. Link, B. N. Narozhny, E. I. Kiselev, and J. Schmalian, Phys. Rev. Lett. 120, 196801 (2018).
	\end{thebibliography}
\end{document}